\documentclass[10pt,aps,pra,twocolumn,floatfix,english,superscriptaddress]{revtex4-1}
\usepackage{amsmath}
\usepackage{amssymb}
\usepackage{graphicx}
\usepackage{braket}
\usepackage[usenames,dvipsnames,svgnames,table]{xcolor}
\usepackage[unicode=true,
            pdfusetitle,
            bookmarks=false,
            bookmarksnumbered=false,
            bookmarksopen=false,
            breaklinks=true,
            pdfborder={0 0 0},
            backref=false,
            colorlinks=true]{hyperref}
\hypersetup{linkcolor=NavyBlue,urlcolor=NavyBlue,citecolor=NavyBlue}

\DeclareMathOperator{\tr}{Tr}
\newcommand{\hp}{\mathsf{H}}

\newcommand{\pemin}{P_{\mathrm{e,\min},M}}
\newcommand{\pemink}{P_{\mathrm{e,\min}|L}}
\newcommand{\diff}{{\rm d}}

\begin{document}

\title{Quantum-optimal detection of one-versus-two incoherent sources with arbitrary separation}

\author{Xiao-Ming Lu}
\affiliation{Department of Electrical and Computer Engineering, National University
of Singapore, 4 Engineering Drive 3, Singapore 117583}

\author{Ranjith Nair}
\affiliation{Department of Electrical and Computer Engineering, National University
of Singapore, 4 Engineering Drive 3, Singapore 117583}

\author{Mankei Tsang}
\affiliation{Department of Electrical and Computer Engineering, National University of Singapore, 4 Engineering Drive 3, Singapore 117583}
\affiliation{Department of Physics, National University of Singapore, 2 Science Drive 3, Singapore 117551}

\date{\today}
\begin{abstract}
  We analyze the fundamental resolution of incoherent optical point
  sources from the perspective of a quantum detection problem:
  deciding whether the optical field on the image plane is generated
  by one source or two weaker sources with arbitrary separation.  We
  investigate the detection performances of two measurement methods
  recently proposed by us to enhance the estimation of the separation.
  For the detection problem, we show that the method of binary
  spatial-mode demultiplexing is quantum-optimal for all values of
  separations, while the method of image-inversion interferometry is
  near-optimal for sub-Rayleigh separations. Unlike the proposal by
  Helstrom, our schemes do not require the separation to be given and
  can offer that information as a bonus in the event of a successful
  detection. For comparison, we also demonstrate the supremacy of our
  schemes over direct imaging for sub-Rayleigh separations.  These
  results demonstrate that simple linear optical measurements can
  offer supremal performances for both detection and estimation.
\end{abstract}

\maketitle

\section{Introduction}
The influential Rayleigh criterion for imaging
resolution~\cite{LordRayleigh1879}, which specifies a minimum
separation between two distinguishable incoherent light sources, is
based on heuristic notions.  A more rigorous approach to the
resolution measure can be formulated in terms of the estimation error
for locating the sources in the presence of
noise~\cite{Ram2006,Chao2016}.  Recently, via quantum estimation
theory~\cite{Helstrom1976,Holevo1982}, it was found that the
estimation of the separation between two incoherent sources below the
Rayleigh criterion can be significantly improved by measurements
employing linear optics and photon
counting~\cite{Tsang2016b,Nair2016,Tsang2016a,Nair2016a,Ang2016,Lupo2016,Rehacek2016,Tsang2016c,Tang2016,Yang2016,Tham2016,Paur2016}.

Besides localization, the resolving power of an imaging system can also be studied via a
detection problem: deciding whether the optical field in the image
plane is generated by one source or two
sources~\cite{Harris1964,Helstrom1973,Acuna1997,Shahram2006,Dutton2010}.
This detection perspective is especially relevant to the detection of
binary stars with telescopes~\cite{Acuna1997} and the detection of
protein multimers with fluorescent microscopes~\cite{Nan2013}. In a
pioneering work, Helstrom obtained the mathematical description of the
quantum-optimal measurement that minimizes the error probability for testing one or two point sources
of thermal light~\cite{Helstrom1973}.  Unfortunately, in addition to having no known physical realization, his method
requires the separation between the two hypothetical sources to be
given, when the separation is usually unknown in practice. A more
recent work by Krovi, Guha, and Shapiro \cite{Krovi2016} investigated the quantum
Chernoff bound \cite{Ogawa2004,Kargin2005,Audenaert2007,Nussbaum2009,Audenaert2008} for this problem and found a measurement scheme that
saturates the bound. In addition, they discovered
that the quantum Chernoff bound can beat the performance of direct
imaging by orders of magnitude. The attainment of this quantum
supremacy without the separation being given remains an open question.

Here we investigate the performance of two practical quantum
measurements for the detection of two weak incoherent point light
sources.  In addition to the error probability, we also assess  the performance of these measurements vis-a-vis direct imaging using the
asymptotic error exponent, which specifies the rate at which the error probability decreases as
the sample size goes to infinity.  We show that a binary spatial-mode
demultiplexing (B-SPADE) scheme~\cite{Tsang2016b} is quantum-optimal
for all values of separations in the following two senses: (1) the
asymptotic error exponent attains the quantum maximum, and (2) the
error probability of a simple decision rule based on the observations
of the B-SPADE can be close to the quantum limit.  We also show that
the scheme of superlocalization by image inversion interferometry
(SLIVER) \cite{Nair2016} is near-optimal for sub-Rayleigh separations. In addition to
the supremacy over direct imaging, our methods do not
require the separation to be given, can offer an accurate estimate of
the separation in the event of a successful detection
\cite{Tsang2016b,Nair2016,Tsang2016a,Nair2016a,Ang2016,Lupo2016,Rehacek2016,Tsang2016c},
and have been experimentally demonstrated in the context of parameter
estimation \cite{Tang2016,Yang2016,Tham2016,Paur2016}. These
additional advantages over the prior proposals by Helstrom
\cite{Helstrom1973} and Krovi \textit{et al.}\ \cite{Krovi2016} hold
tremendous promise for practical detection applications in both
astronomy \cite{Acuna1997} and molecular imaging \cite{Nan2013}.

The paper is organized as follows.  In Sec.~\ref{sec:model}, we
introduce the one-source-versus-two hypothesis testing problem along with our model of the sources and the imaging system.  In Sec.~\ref{sec:hypothesis_tesing}, we
calculate the minimum probability of error and derive the optimal
measurement.  In Sec.~\ref{sec:practical_measurement}, we investigate
the performances of the SPADE and SLIVER schemes in terms of their
asymptotic error exponents and average error probabilities with 
simplified decision rules.  In Sec.~\ref{sec:performance}, we explicitly compare
the performance of the B-SPADE and SLIVER schemes with  conventional direct imaging for the case of a Gaussian point-spread
function.  We summarize our results in Sec.~\ref{sec:conclusion}.

\section{One source versus two sources}\label{sec:model}

\begin{figure}[tb]
	\centering
	\includegraphics[width=1\linewidth]{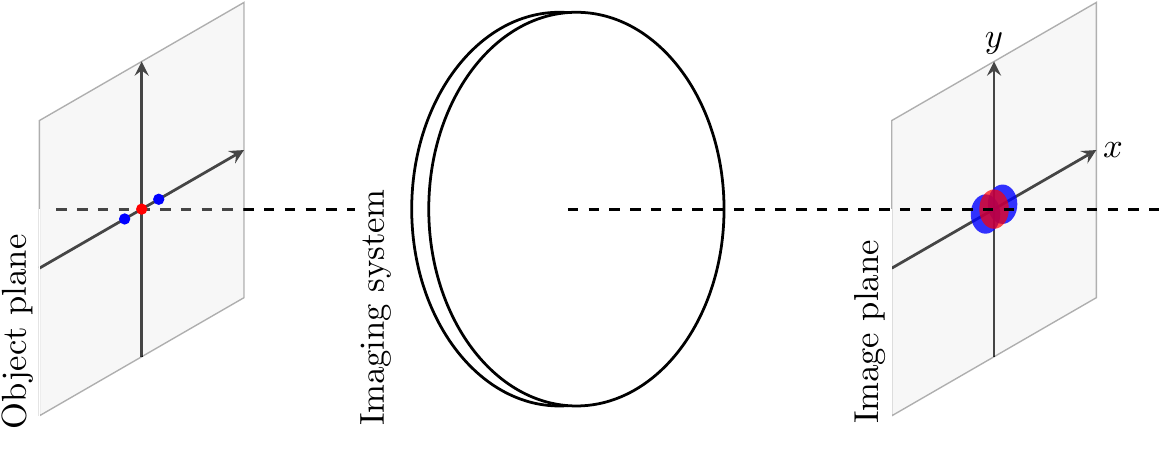}
	\caption{
	Illustration of the imaging of incoherent point light sources. 
	The images of two closely spaced point sources with the same intensity (blue) appear as that of one point source located midway with twice intensity (red).
	}
	\label{fig:model}
\end{figure}

Denote by $\hp_2$ the hypothesis that there are two point sources
emitting photons with equal intensities, and $\hp_1$ the hypothesis
that there is only one point source located midway with twice the
intensity, see Fig.~\ref{fig:model}.  Let $\rho_1$ and $\rho_2$ be the
density operators for the quantum optical fields arriving at the image plane
under $\hp_1$ and $\hp_2$, respectively.  Assuming that the point
sources of light are weak and incoherent, the density operators per
temporal mode can be approximated as~\cite{Tsang2016b}
\begin{equation}\label{eq:rho}
	\rho_i = (1-\epsilon)\ket{\rm vac} \bra{\rm vac} + \epsilon \eta_i
\end{equation}
for $i=1,2$, where $\epsilon\ll1$ is the average photon number
arriving on the image plane, $\ket{\rm vac}$ denotes the vacuum state,
$\eta_{i}$ are the corresponding one-photon states, and
$O(\epsilon^2)$ terms have been neglected
\cite{Tsang2016b,Gottesman2012,Tsang2011}. This approximation enables
us to simplify the theory in comparison with
Refs.~\cite{Helstrom1973,Krovi2016} and still obtain similar results.
We assume that the imaging system is spatially invariant and the two
hypothetic point sources lie along the $x$-axis.  Then, the one-photon
states under the two hypotheses can be expressed as
\begin{align}
	\eta_1 &= \ket{\psi_1}\bra{\psi_1},
	\label{eq:eta1}\\
	\eta_2 &= \frac12\ket{\psi_+}\bra{\psi_+} + \frac12\ket{\psi_-}\bra{\psi_-},
	\label{eq:eta2} \\
	\ket{\psi_1}   &\equiv \int_{-\infty}^\infty\diff x \int_{-\infty}^\infty\diff y\,\psi(x,y)\ket{x,y},
	\label{eq:psi1}\\
	\ket{\psi_\pm} &\equiv \int_{-\infty}^\infty\diff x \int_{-\infty}^\infty\diff y\,\psi(x\pm d/2,y) \ket{x,y},
	\label{eq:psi2}
\end{align}
where $(x,y)$ is the image-plane coordinate normalized with respect to the magnification factor of the imaging system, $\psi(x,y)$ is the point-spread function for the imaging system, and $d$ is the separation between the hypothetical two point sources.

\section{Hypothesis testing}\label{sec:hypothesis_tesing}

A strategy for accepting one or the other hypothesis, known as a decision rule, is
given by partitioning the space of observations into two regions,
denoted by $\Omega_1$ and $\Omega_2$: The one-source hypothesis
$\hp_1$ is accepted if the observation belongs to $\Omega_1$, and the
two-source hypothesis $\hp_2$ is accepted otherwise.  The performed
quantum measurement can be described by a positive-operator-valued
measure (POVM) $\{E(z)\}$, where $z$ denotes the outcome, and $E(z)$'s
are nonnegative operators resolving the identity operator as
$\int\!{\rm d}\mu(z) E(z)=\openone$ with $\mu(z)$ being an appropriate
measure of $z$ \cite{Helstrom1976,Holevo1982,Hayashi2006}.  Defining $E_1=\int_{z\in\Omega_1}\!\diff\mu(z) E(z)$
and $E_2=\int_{z\in\Omega_2}\!\diff\mu(z) E(z)$, the probabilities of
the type-I (false-alarm) and type-II (miss) errors for a
one-source-versus-two testing are given by
\begin{equation}
	\alpha \equiv \tr(E_2\rho_1^{\otimes M}) 
	\quad\mbox{and}\quad 
	\beta \equiv \tr(E_1\rho_2^{\otimes M}),
\end{equation}
respectively, where $M$ is the number of available temporal modes.
This number $M$ is also referred to as the sample size.
Assuming prior probabilities $p_1$ and $p_2$ for the respective hypotheses, the average probability of errors is
$P_{\mathrm{e},M} \equiv p_1 \alpha + p_2 \beta$, 
which is taken as the figure of merit of a measurement and decision strategy~\cite{Helstrom1976}.
The minimum error probability optimized over all quantum measurements and classical decision rules is given by~\cite{Helstrom1976}
\begin{equation}\label{eq:pe_min}
	\pemin =\frac{1}{2}(1-\Vert p_2\rho_2^{\otimes M} - p_1\rho_1^{\otimes M}\Vert_1),
\end{equation}
where $\Vert A \Vert_{1} \equiv \tr\sqrt{A^\dagger A}$ denotes the trace norm. 
Moreover, the minimum error probability can be achieved by the Helstrom-Holevo test for which $E_2$ is the projector onto the eigen subspace of $p_2\rho_2^{\otimes M} - p_1\rho_1^{\otimes M}$ with positive eigenvalues~\cite{Helstrom1976,Holevo1973b}.

\subsection{Minimum error probability}

For the model described by Eqs.~(\ref{eq:rho}--\ref{eq:eta2}), the minimum error probability is given by
\begin{align}
	\pemin &= \sum_{L=0}^{M} \binom{M}{L}(1-\epsilon)^{M-L}\epsilon^L \pemink, 
	\label{eq:binomial}\\
	\pemink &\equiv \frac12-\frac12\Vert p_2\eta_2^{\otimes L} - p_1\eta_1^{\otimes L}\Vert_1.
	\label{eq:pe_conditioned}
\end{align}
Notice that the $\binom{M}{L}(1-\epsilon)^{M-L}\epsilon^L$ is the
probability of $L$ photons arriving at the imaging plane, and $\pemink$
is the minimum probability of error conditioned on knowing that $L$
photons are detected on the imaging plane.  The form of Eq.~\eqref{eq:binomial}
is due to the fact that the distinguishability between $\rho_1$ and
$\rho_2$ lies in the one-photon sector and the zero-photon
event is uninformative.  Either the conditional error probability or
the unconditional one can be used as a figure of merit, depending on
whether or not the number of the photons arriving at the image plane is measured.

We assume that the inner products $\braket{\psi_\pm|\psi_1}$ and $\braket{\psi_+|\psi_-}$ are real, which is satisfied when the point-spread function $\psi(x,y)$ in Eqs.~\eqref{eq:psi1} and \eqref{eq:psi2} is symmetric about $y$-axis, namely, $\psi(x,y)=\psi(-x,y)$.
We then obtain
\begin{equation}\label{eq:analytical}
	\pemink  = \frac{p_1 + p_2\lambda_+^L}{2} \left[
	1 - \sqrt{1-\frac{4p_1p_2\chi^{2L}}{(p_1+p_2\lambda_+^L)^2}}
	\right],
\end{equation}
where 
\begin{align} 
\lambda_+ &= (1+\braket{\psi_+|\psi_-})/2 \equiv [1+\delta(d)]/2, \label{eq:lambda+}\\
\chi &=\braket{\psi_+|\psi_1} \equiv \delta(d/2), \label{eq:chi}
\end{align}
where we have introduced the (real-valued) overlap function 
\begin{equation}
\delta(d)= \int\!\diff x\diff y\,\psi^*(x,y)\psi(x-d,y)
\end{equation}
corresponding to the given point-spread function.

As long as the number of detected photons is sufficiently large such that $L\gg\log(p_1/p_2)/(2\log\chi)$, the minimum error probability Eq.~\eqref{eq:analytical} can be approximated to 
\begin{equation}\label{eq:approx_pe}
 	\pemink\approx\frac{p_1p_2\chi^{2L}}{p_1+p_2\lambda_+^L}\lesssim p_2\chi^{2L}.
\end{equation}
Considering that a decision is usually made after a sufficient number of photons are collected, the above approximation would be much useful in realistic situations.

To derive Eq.~\eqref{eq:analytical}, we first diagonalize the density operator $\eta_2$ as
$\eta_2 = \lambda_+ \ket{\phi_+}\bra{\phi_+} + \lambda_- \ket{\phi_-}\bra{\phi_-}$,
where $\lambda_-\equiv[1-\delta(d)]/2$ and the eigenstates are given by
\begin{align}
	\ket{\phi_\pm} \equiv \frac{1}{2\sqrt{\lambda_\pm}}(\ket{\psi_+} \pm \ket{\psi_-}).
\end{align}
The pure states $\ket{\phi_+}$ and $\ket{\phi_-}$ are symmetric and antisymmetric about $y$-axis, respectively. 
Note that the antisymmetric eigenstate $\ket{\phi_-}$ is orthogonal to $\ket{\psi_1}$, leading to 
\begin{align}\label{eq:decompose}
	\Vert p_2\eta_2^{\otimes L} - p_1\eta_1^{\otimes L}\Vert_1
	= \Vert \Gamma \Vert_1 + p_2(1 - \lambda_+^L)
\end{align}
with $\Gamma \equiv 
	p_2\lambda_+^L (\ket{\phi_+}\bra{\phi_+})^{\otimes L} - 
	p_1(\ket{\psi_1}\bra{\psi_1})^{\otimes L}. 	
$
By diagonalizing $\Gamma$ in the two-dimensional subspace spanned by $\ket{\phi_+}^{\otimes L}$ and $\ket{\psi_1}^{\otimes L}$, we obtain
\begin{align}
	\Vert\Gamma\Vert_1
 	& = \sqrt{(p_1+p_2\lambda_+^L)^2 - 4p_1p_2\chi^{2L}},
\end{align}
which together with Eq.~\eqref{eq:decompose} leads to Eq.~\eqref{eq:analytical}.

\subsection{Optimal measurement}

The optimal measurement is suggested by the derivation of the minimum error probability as follows.
First, since the two-source antisymmetric state $\ket{\phi_-}$ is orthogonal to the one-source state $\ket{\psi_1}$, one can infer that there are definitely two point sources if any population in $\ket{\phi_-}$ is observed.
It is therefore essential to include $\ket{\phi_-}$ in the measurement basis.
Secondly, in order to achieve the minimum error probability, the measurement is further required to distinguish $\ket{\phi_+}^{\otimes L}$ and $\ket{\psi_1}^{\otimes L}$, whose updated probabilities are $p_2\lambda_+^L$ and $p_1$ respectively.
According to the Helstrom-Holevo theorem~\cite{Helstrom1976,Holevo1973b}, the optimal measurement can be given by including the eigenstates of $\Gamma$ with nonzero eigenvalues in the measurement basis.
The complete Helstrom-Holevo test can be described by the following POVM element associated to $\hp_1$:
\begin{align}
	E_1 &= (\openone-\ket{\phi_-}\bra{\phi_-})^{\otimes L} - \ket\Psi\bra\Psi,
\end{align}
where $\ket\Psi$ is the eigenstate of $\Gamma$ with the positive eigenvalue.

However, the quantum measurement underlying the Helstrom-Holevo test is difficult to be physically implemented from the following aspects.
First, the optimization is over all possible quantum measurements, and the optimal measurement is a joint one over multiple samples~\cite{Hayashi2006}.
Secondly, the optimal measurement in general depends on the separation between the two hypothetic point sources, which is often unknown in the first place.
Lastly, the optimal measurement in general depends on the ratio of the prior probabilities of the two hypotheses, whose determination is often subjective.

\section{Practical measurements}\label{sec:practical_measurement}

To circumvent the difficulties in the Helstrom-Holevo test, we
consider the measurements of B-SPADE~\cite{Tsang2016b} and
SLIVER~\cite{Nair2016}.  These two methods are recently shown to be
good at estimating the separation between two closely-spaced
incoherent point sources, for which direct imaging performs poorly.

In addition to the error probability, we also use the asymptotic error exponent to assess the performance of quantum measurements for  one-source-versus-two hypothesis testing.
For a specific quantum measurement performed on each sample, it is known that the minimum error probability over all decision rules decreases exponentially in $L$ as $\pemink^{\rm (meas)}\sim\exp(-L\xi^{\rm (meas)})$ for a large number $L$ of detected photons~\cite{chernoff1952,VanTrees2013,Cover2012}, where `meas' labels the measurement.
The asymptotic error exponent (conditioned on the presence of a photon) is given by the Chernoff distance (also known as Chernoff information or Chernoff exponent)~\cite{chernoff1952,VanTrees2013,Cover2012}:
\begin{equation}\label{eq:classical_Chernoff}
	\xi^{(\textrm{meas})} = -\log\min_{0\leq s\leq1} \int\!\mathrm{d} \mu(z)\, \Lambda_1(z)^s \Lambda_2(z)^{1-s},
\end{equation}
where $\Lambda_j(z)=\tr[E(z)\eta_j]$ is the probability of obtaining the outcome $z$ under the hypothesis $\hp_j$, and $\{E(z)\}$ is the POVM for the measurement.
The asymptotic error exponent is bounded from above by the so-called quantum Chernoff distance as~\cite{Ogawa2004,Kargin2005,Audenaert2007,Nussbaum2009,Audenaert2008}: 
\begin{align}\label{eq:chernoff}
	\xi^{\rm (meas)}\leq \xi \equiv -\log\min_{0\leq s\leq1}\tr(\eta_1^s\eta_2^{1-s})
		= -2\log\chi.
\end{align}

\subsection{B-SPADE}

Spatial-mode demultiplexing is a measurement method that demultiplexes the image-plane optical field into the desired orthogonal spatial modes in which the photon is detected~\cite{Tsang2016b}. 
The binary version of spatial-mode demultiplexing---the B-SPADE---only distinguishes photons in a specific mode and all other modes.
Consider such a B-SPADE scheme that distinguishes the mode $\ket{\psi_1}$ and its orthogonal-complement modes.
In other words, the POVM for the measurement on each sample is $\{\ket{\psi_1}\bra{\psi_1},\openone - \ket{\psi_1}\bra{\psi_1}\}$. 
In such a case, it is easy to see that the asymptotic error exponent for this B-SPADE measurement is 
\begin{equation}
	\xi^{\rm (B-SPADE)} = -2\log\chi.
\end{equation}
This attains the quantum limit given by Eq.~\eqref{eq:chernoff}, meaning that the B-SPADE measurement has the optimal asymptotic error exponent.

To make a decision on hypotheses based on the observations of the B-SPADE, one needs to identify a decision rule.
For a given separation, the optimal decision rule is given by the likelihood-ratio test~\cite{VanTrees2013}: 
For a given observation data $(z_1,z_2,\ldots,z_L)$, choose $\hp_2$ if $\prod_{j=1}^L\Lambda_2(z_j)/\Lambda_1(z_j)>p_1/p_2$, and choose $\hp_1$ otherwise.
If the separation is unknown, one can use the generalized-likelihood-ratio test~\cite{Kay1998}, which first estimates the separation and then does the likelihood-ratio test with the estimated value.
Here, we give a simplified decision rule that is irrelevant to the separation as follows.
If any detectors associated to the POVM element $\openone-\ket{\psi_1}\bra{\psi_1}$ click during the observation, one can infer that there is definitely two point sources, i.e., $\hp_2$ is true.
The simplified decision rule is given by just accepting $\hp_1$ if all the detectors click during the observation are associated to the mode $\ket{\psi_1}$. 
This decision rule agrees with the likelihood-ratio test when $\chi^{2L}\leq p_1/p_2$, which is always true for the cases of $p_1\geq p_2$.

With the simplified decision rule, the POVM element associated to $\hp_1$ is $E_1=(\openone-\ket{\phi_-}\bra{\phi_-})^{\otimes L}$ on the Hilbert space of $L$ detected photons.
It can be shown that the error probabilities are
\begin{align}
	\alpha^{\rm (B-SPADE)} = 0,
	\quad
	\beta^{\rm (B-SPADE)} = \chi^{2L},
\end{align} 
and the (unconditional) error probability is
\begin{equation}
	P_{\mathrm{e},M}^\mathrm{(B-SPADE)} = p_2(1-\epsilon+\epsilon\chi^2)^M
\end{equation}
for $M$ temporal modes.

Comparing with Eq.~\eqref{eq:approx_pe}, it can be seen that conditional probability of errors given $L$ detected photons of the B-SPADE scheme with the simplified decision rule is close to its minimum, when the conditions $\chi^{2L}\ll p_1/p_2$ and $\lambda_+^L\ll p_1/p_2$ are satisfied.
Since $\chi\in(-1,1)$ and $\lambda_+\in(0,1)$, these two conditions can always be satisfied for a sufficiently large $L$, meaning that the scheme considered here is asymptotically optimal.

\subsection{SLIVER}
The second practical measurement we consider is  SLIVER, which separates the optical field on the image plane into the symmetric and antisymmetric components with respect to the inversion at the origin, and then detects photons in the respective ports~\cite{Nair2016}.
Here, we consider a modified SLIVER for which the inversion operation is replaced by the reflection operation with respect to $y$-axis -- this modification corresponds to the Pix-SLIVER scheme of \cite{Nair2016a} with single-pixel (bucket) detectors at the two outputs.
For simplicity, we just refer to this modified version as  SLIVER henceforth.

Notice that the states $\ket{\psi_1}$ and $\ket{\phi_+}$ are symmetric and $\ket{\phi_-}$ is antisymmetric under the reflection with respect to the $y$-axis.
Thus, a detected photon under the hypothesis $\hp_2$ is at the symmetric port with probability $\lambda_+$ and at the antisymmetric port with probability $\lambda_-$, while a detected photon under $\hp_1$ is always at the symmetric port.
According to Eq.~\eqref{eq:classical_Chernoff}, we obtain the asymptotic error exponent
\begin{equation}\label{eq:chernoff_sliver}
	\xi^{\rm (SLIVER)} = -\log\lambda_+.
\end{equation}

We also give a simplified decision rule similar to that for the B-SPADE scheme: 
The hypothesis $\hp_2$ is accepted if any photon is detected at the antisymmetric port during the whole observation, and $\hp_1$ is accepted otherwise.  
This decision rule agrees with the likelihood-ratio test when $\lambda_+^L \leq p_1/p_2$, which is always true for the cases of $p_1\geq p_2$.
In such a case, the probabilities of the type-I and type-II errors are given by
\begin{equation}
	\alpha^{\rm (SLIVER)} = 0, \quad \beta^{\rm (SLIVER)} = \lambda_+^L,
\end{equation}
respectively, leading to the (unconditional) error probability
\begin{equation}
	P_{\mathrm{e},M}^\mathrm{(SLIVER)} =p_2(1-\epsilon+\epsilon\lambda_+)^M
\end{equation}
for $M$ temporal modes.

\section{Performance comparison}\label{sec:performance}

\begin{figure}[tb]
	\includegraphics[]{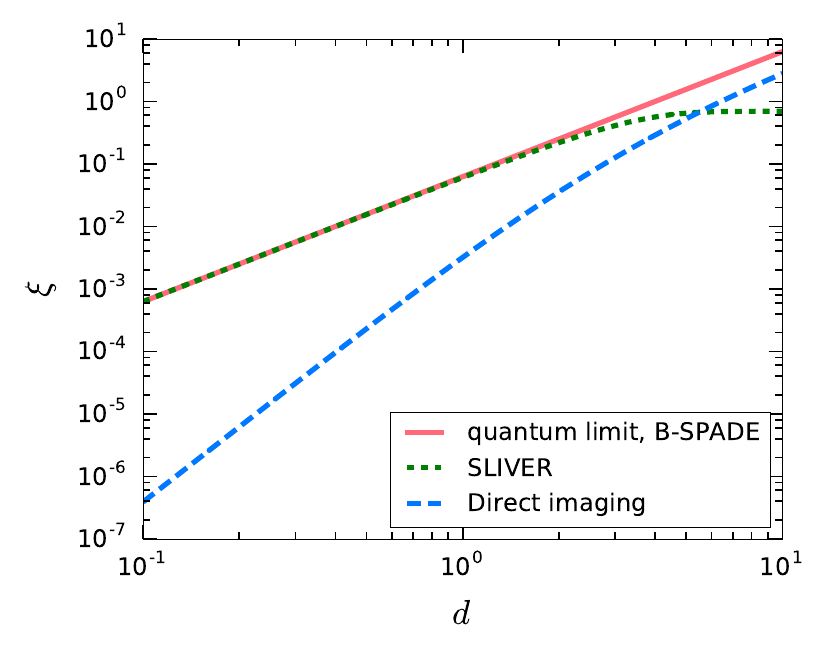}
	\caption{\label{fig:chernoff} 
	Asymptotic error exponents versus the separation between the two hypothetic incoherent point sources, for the B-SPADE scheme (red solid), the SLIVER scheme (green dotted), and the direct imaging (blue dash).
	The asymptotic error exponents of the B-SPADE scheme attains its quantum limit given by the quantum Chernoff distance for all values of $d$.
	}
\end{figure}

\begin{table*}[tb]
    \caption{\label{tab}Asymptotic error exponents and average error probabilities: Expressions in terms of $\chi$ and $\lambda_+$ (given by Eqs.~\eqref{eq:lambda+}-\eqref{eq:chi}) are valid for general symmetric point-spread functions.  The asymptotic error exponents are given explicitly for the case of the Gaussian point-spread function of Eq.~\eqref{eq:gaussian} using Eq.~\eqref{eq:gaussian_result}. For the B-SPADE and SLIVER scheme, we use the simplified decision rules.}
    \begin{ruledtabular}
        \begin{tabular}{llll}
        Scheme & Asymptotic error exponent & Conditional error probability & Unconditional error probability \tabularnewline
        \hline
        	Quantum limit & $d^2/16$ & Eq.~\eqref{eq:analytical}, $ \approx p_2\chi^{2L}$ for $L \gg1$ & $p_2(1-\epsilon+\epsilon\chi^2)^M$ for a large $N =M\epsilon$ \tabularnewline
            B-SPADE & $d^2/16$ & $p_2\chi^{2L}$ & $p_2(1-\epsilon+\epsilon\chi^2)^M$ \tabularnewline
            SLIVER & $d^2/16 - d^4/512 + O(d^6)$ & $p_2\lambda_+^L$ & $p_2(1-\epsilon+\epsilon\lambda_+)^M$ \tabularnewline
            Direct imaging & $d^4/256+O(d^6)$ & bounds given by Eq.~\eqref{eq:bounds} & bounds given by Eqs.~\eqref{eq:binomial} and \eqref{eq:bounds}\tabularnewline
        \end{tabular}
	\end{ruledtabular}
\end{table*}

Here, we compare the performance of the B-SPADE scheme, the SLIVER scheme, and the conventional measurement---direct imaging.
We assume that the point-spread function is Gaussian as $\psi(x,y)=f(x)f(y)$ with
\begin{equation}\label{eq:gaussian}
	f(x) = \frac{1}{(2\pi\sigma^2)^{1/4}}\exp\left(-\frac{x^2}{4\sigma^2}\right),\quad\sigma=\frac{\lambda}{2\pi\mathrm{NA}},
\end{equation}
where $\lambda$ is the free-space wavelength and NA is the effective numerical aperture.
For simplicity, we rescale the $x$ coordinate by taking $\sigma$ as the unit of length in what follows.
Consequently, $d$ is defined as a dimensionless number in units of $\sigma$.
In such a case, it can be shown that
\begin{align}\label{eq:gaussian_result}
	\lambda_+ = \frac12+\frac12\exp\left(-\frac{d^{2}}{8}\right),
	\quad
	\chi  = \exp\left(-\frac{d^{2}}{32}\right),
\end{align}
with which one has the asymptotic error exponents and the error probabilities of the B-SPADE and SLIVER schemes as well as their quantum limits.
We list the relevant results in Table~\ref{tab} for convenience.



The asymptotic error exponent of the B-SPADE is $d^2/16$, which is the same as its quantum limit.
For the SLIVER measurement, applying the Taylor series expansion with respect to $d$ on Eq.~\eqref{eq:chernoff_sliver} with Eq.~\eqref{eq:gaussian_result} yields 
\begin{equation}
	\xi^{\rm (SLIVER)} =\frac{d^2}{16} - \frac{d^4}{512}+O(d^6).
\end{equation}
It can be seen that the error exponent of the SLIVER scheme is close to the quantum limit when the separation is small, as shown in Fig.~\ref{fig:chernoff}.

\begin{figure}[tb]
	\includegraphics[]{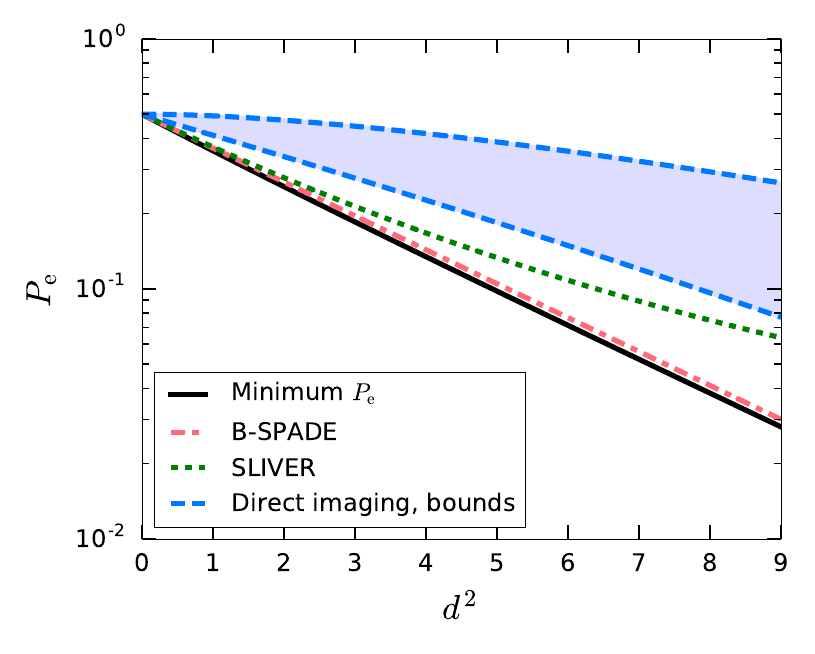}
	\caption{\label{fig:nonasym} 
	Conditional error probabilities and their bounds as functions of the separation squared, conditioned on $L=5$ detected photons.
	The prior probabilities of the two hypotheses are assumed to be equal.
	For the B-SPADE and SLIVER schemes, we use the simplified decision rules elucidated in the main text.
	The minimal error probability of the direct imaging must lie in the shaded region described by Eq.~\eqref{eq:bounds}.
	}
\end{figure}

\begin{figure}[tb]
	\centering
	\includegraphics[]{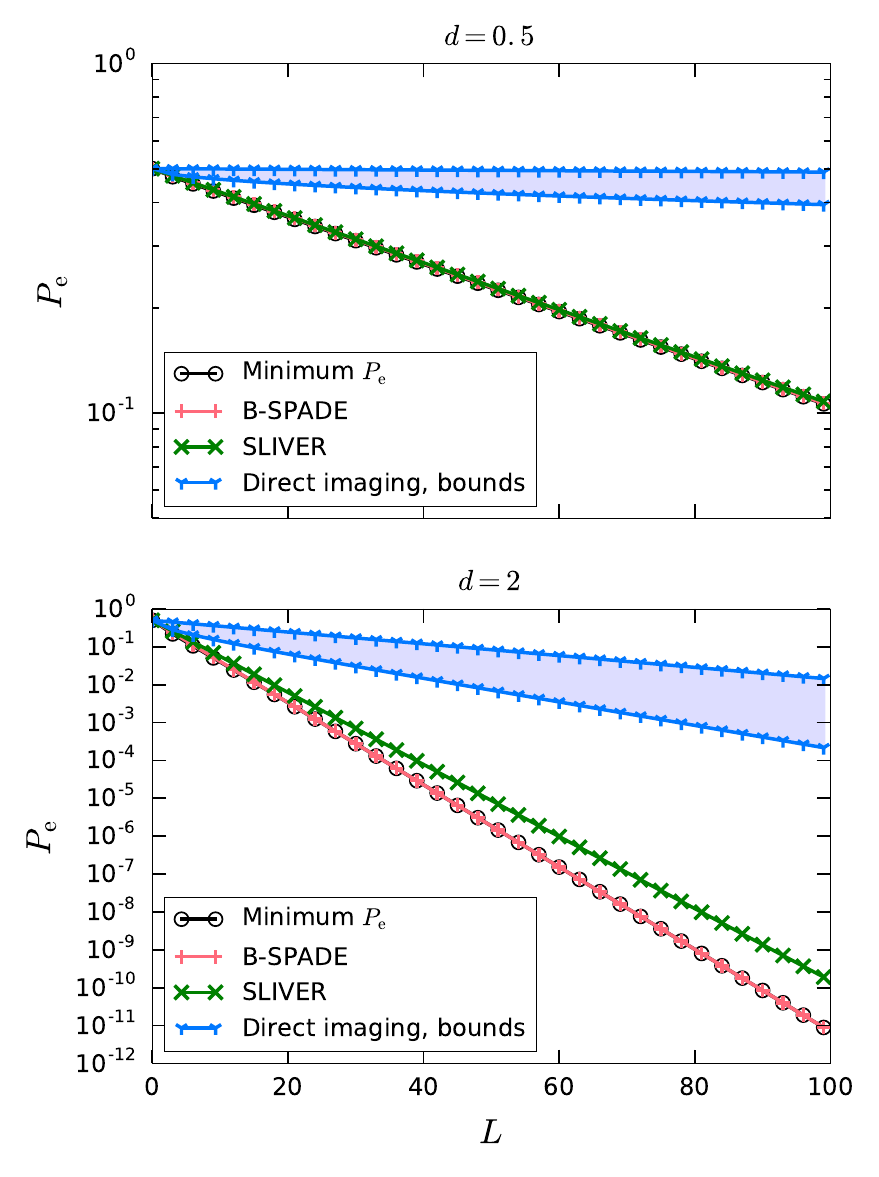}
	\caption{Average probabilities of errors conditioned on $L$ detected photons for $d=0.5$ (the upper plot) and $d=2$ (the lower plot).
	The prior probabilities of the two hypotheses are assumed to be equal.
	The minimal error probability of the direct imaging must lie in the shaded region described by Eq.~\eqref{eq:bounds}.
	The lines are guides for eyes.}
	\label{fig:pe}
\end{figure}

To elucidate the necessity of the new technologies like the B-SPADE and SLIVER schemes for improving the imaging resolution based on the one-source-versus-two hypothesis testing, we also calculate the performance of direct imaging.
Conditioned on the presence of a photon, direct imaging of the image-plane field using an ideal continuum photodetector results in the probability densities
\begin{align}
\Lambda_1(x,y) &= \left| \psi(x,y)\right|^2,\\
\Lambda_2(x,y) &= \frac{1}{2} \left| \psi(x-d/2,y)\right|^2 +  \frac{1}{2} \left| \psi(x+d/2,y)\right|^2 
\end{align}
for the position of arrival of the photon under  $\textsf{H}_1$ and $ \textsf{H}_2$ respectively.
For the Gaussian point-spread function, the error exponent is given by
\begin{align}
	\xi^{({\rm direct})} 
 	& = -\log \min_{0\leq s\leq 1} \zeta(s), \\
 	\zeta(s) &\equiv \exp\left(-\frac{sd^2}{8}\right)
 	\mathbb{E}\left[\cosh^{s}\left(\frac{xd}{2}\right)\right], \label{eq:zeta_direct}
\end{align}
where $\mathbb{E}[\bullet]\equiv \int_{-\infty}^\infty{\rm d}x\,\bullet \exp(-x^2/2)/(\sqrt{2\pi})$ denotes the expectation value with respect to a Gaussian distribution of $x$ with zero mean and unit variance.
Using the Taylor series expansion with respect to $d$ and the moments $\mathbb{E}[x^2] = 1$ and $\mathbb{E}[x^4] = 3$, we obtain 
\begin{align}\label{eq:taylor}
	\zeta(s) = 1 - \frac{s(1-s)d^4}{64} + O(d^6).
\end{align}
For small separations, $\zeta(s)$ takes its minimum at $s\approx1/2$, leading to
\begin{equation}\label{eq:approximation}
	\xi^{\rm (direct)} \approx -\log \left(1 - \frac{d^4}{256}\right) \approx \frac{d^4}{256}.
\end{equation}
This approximation, as well as the result of numerically optimizing $\xi^{\rm (direct)}$, is shown in Fig.~\ref{fig:chernoff}, implying that direct imaging performs poorly for the  one-source-versus-two testing for small separations.

It is difficult to obtain the analytic result for the minimal error probability of direct imaging. 
We here resort to the following upper and lower bounds in the case of equal priors ($p_1=p_2=1/2$)~\cite{Fuchs1999,Cover2012}
\begin{equation}\label{eq:bounds}
 	\frac{1}{2}(1-\sqrt{1-F^{2L}})
 	\leq\pemink^{(\textrm{direct})} 
 	\leq \frac{1}{2} \exp(-L\xi^{(\mathrm{direct})}),
\end{equation} 
where $F \equiv \int\!\diff x\diff y\sqrt{\Lambda_1(x,y)\Lambda_2(x,y)}$ is the Bhattacharyya coefficient for the two  probability distributions generated by direct imaging.
For small separations, by noting $F=\zeta(1/2)$ and that $\xi^\mathrm{(direct)}\approx-\log\zeta(1/2)$ as shown above, we obtain $F\approx\exp(-\xi^\textrm{(direct)})$.
The upper and lower bounds Eq.~\eqref{eq:bounds} are used in Figs.~\ref{fig:nonasym} and \ref{fig:pe}.

Figure~\ref{fig:chernoff} plots the asymptotic error exponents of different measurement schemes.
It can be seen that the B-SPADE measurement has the optimal asymptotic error exponent for all values of the separation.
For the SLIVER measurement, the asymptotic error exponent is close to the quantum limit for small separations, but diverges from the quantum limit  when the separation increases, and becomes worse than that of direct imaging for sufficiently large separations $d \gtrsim 5$.
The asymptotic error exponent of direct imaging is much smaller than the quantum limit for small separations, meaning that direct imaging is asymptotically inefficient for small separations.

The asymptotic error exponent only reflects the performance in the situations where the number of detected photons is sufficiently large.
For a small number of detected photons like $L=5$, we plot in Fig.~\ref{fig:nonasym} the average conditional probabilities of errors for different schemes in the case of equal priors. 
The relation to any $L$ is plotted in Fig.~\ref{fig:pe} for the cases of $d=0.5$ and $d=2$.
It can seen from Figs.~\ref{fig:nonasym} and \ref{fig:pe} that the performance profile we have obtained in terms of the asymptotic error exponents still holds, at least for the case of equal priors, when the number of detected photons is small:
The B-SPADE scheme is near-optimal, the SLIVER scheme has a good performance only for small separations, whereas direct imaging performs poorly.

\begin{figure}[tb]
	\centering
	\includegraphics{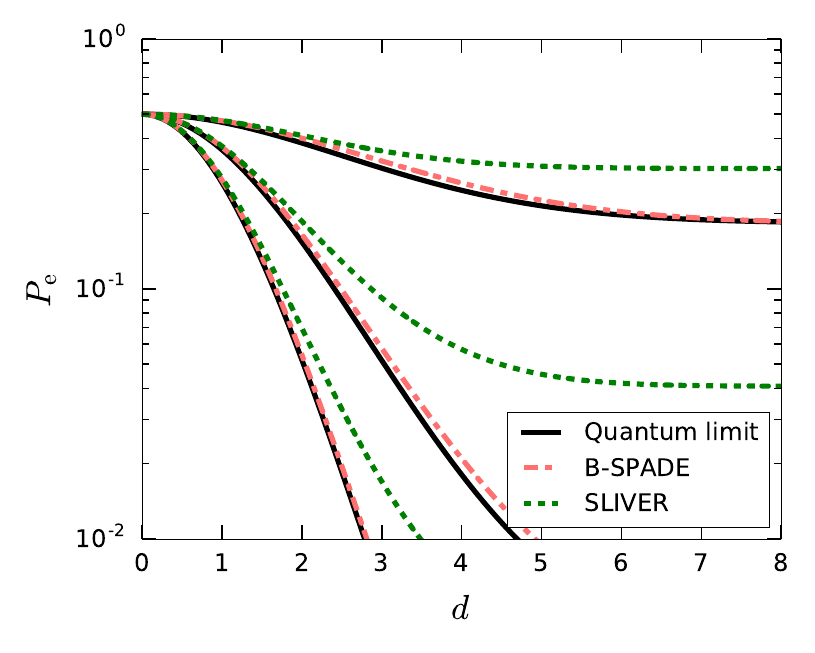}
	\caption{Unconditional error probability in the case of equal priors and the Gaussian point-spread function for $M=100, 500,$ and $1000$ (from above to below). Here we set $\epsilon=0.01$.}
	\label{fig:unconditional}
\end{figure}

In fact, a single on-off detector, which resolves neither the arrival
time nor the photon number, is sufficient for the simplified decision
rules used in this work.  This on-off detector can be associated to
the orthogonal complement of the mode $|\psi_1\rangle$ for the B-SPADE
and to the antisymmetric components for the SLIVER.   Hypothesis $\hp_2$ is accepted if and only if it clicks.  If we need to simultaneously know
the conditional error probability, then at least two
photon-number-resolving detectors are required such that the total
number of the photons arriving on the image plane can be obtained from
the observation.  Alternatively, the (unconditional) error probability
may be used to quantify the performance of a testing scheme, as shown
in Fig.~\ref{fig:unconditional}.  In such a case, one needs the
average photon number per temporal mode, $\epsilon$, and the number of
temporal modes, $M$, rather than the total number of photons arriving
in imaging plane.

\section{Conclusion}\label{sec:conclusion}

We have analyzed the imaging resolution problem from the perspective
of quantum detection theory.  In terms of the asymptotic error
exponents (regardless of the priors) and the average error
probabilities (in the case of equal priors), we have shown that the
B-SPADE and SLIVER measurements are superior to direct imaging for
resolving two close incoherent point sources.  Compared with the prior
proposals by Helstrom \cite{Helstrom1973} and Krovi \textit{et al.}\
\cite{Krovi2016}, our measurements do not need to know the separation
between the two hypothetical sources and can offer that information as
a bonus
\cite{Tsang2016b,Nair2016,Tsang2016a,Nair2016a,Ang2016,Lupo2016,Rehacek2016,Tsang2016c,Tang2016,Yang2016,Tham2016,Paur2016}
in the event of a successful detection. 
Furthermore, we have proposed simple decision rules, which are independent of the two-source separation and the prior probabilities of the hypotheses, and can be implemented using a single on-off detector without temporal resolution over the observation interval.  
Given the rapid recent experimental progress
\cite{Tang2016,Yang2016,Tham2016,Paur2016}, applications of our
techniques to astronomy and molecular imaging analysis may be expected
in the near future.

\acknowledgments M.~T.\ acknowledges inspiring discussions with Saikat
Guha, who informed us of their results \cite{Krovi2016} as early as 2012
and motivated us to look into the detection problem. M.~T.\ also
thanks Geoff Stiebinger, who informed us of the relevance of the
detection problem to protein multimer analysis \cite{Nan2013}.  This
work is supported by the Singapore National Research Foundation under
NRF Grant No.~NRF-NRFF2011-07 and the Singapore Ministry of Education
Academic Research Fund Tier 1 Project R-263-000-C06-112.

\bibliography{onevstwo}

\end{document}